\documentclass[twocolumn,preprintnumbers,amsmath,amssymb]{revtex4-1}

\usepackage{graphicx}
\usepackage{dcolumn}
\usepackage{bm,epsfig}
\usepackage{epstopdf}

\begin{document}

\title{Nematic state in CeAuSb$_{2}$}

\author{S. Seo$^{1}$, Xiaoyu Wang$^{2}$, S. M. Thomas$^{1}$, M. C. Rahn$^{1}$, D. Carmo$^{3}$, F. Ronning$^{1}$, E. D. Bauer$^{1}$, R. D. dos Reis$^{3}$, M. Janoschek$^{1,4}$, J. D. Thompson$^{1}$, R. M. Fernandes$^{5}$ and P. F. S. Rosa$^{1}$}
\affiliation{
$^{1}$ Los Alamos National Laboratory - Los Alamos, NM 87545, USA\\
$^{2}$ James Frank Institute - University of Chicago - Chicago, IL 60615, USA\\
$^{3}$ Brazilian Synchrotron Light Laboratory (LNLS) - Campinas - Sao Paulo, Brazil. \\
$^{4}$ Laboratory for Scientific Developments and Novel Materials - Paul Scherrer Institut - Villigen PSI - Switzerland. \\
$^{5}$ School of Physics and Astronomy - University of Minnesota  - Minneapolis MN 55455, USA \\}

\begin{abstract}
At ambient pressure and zero field, tetragonal CeAuSb$_{2}$ hosts stripe antiferromagnetic order at $T_{N} = 6.3$~K. Here we first show \textit{via} bulk thermodynamic probes and x-ray diffraction measurements that this magnetic order is connected with a structural phase transition to a superstructure which likely breaks $C_{4}$ symmetry, thus signaling nematic order. The temperature-field-pressure phase diagram of CeAuSb$_{2}$ subsequently reveals the emergence of additional ordered states under applied pressure at a multicritical point. Our phenomenological model supports the presence of a vestigial nematic phase in CeAuSb$_{2}$ akin to iron-based high-temperature superconductors; however, superconductivity, if present, remains to be discovered. 
 \end{abstract}

\maketitle

\section{INTRODUCTION}

Heavy-fermion systems, which consist of a lattice of $f$-electrons that hybridize with the conduction electron sea, host prototypical examples of strongly correlated electronic states
 \cite{Stewart1984,Fisk1988,Coleman2007,Coleman2012}. In particular, tetragonal Ce-based compounds often reveal novel quantum states of matter in the vicinity of a 
 quantum critical point (QCP) at which a magnetic transition is suppressed to zero temperature by non-thermal parameters, e.g. pressure, magnetic field, or chemical doping 
 \cite{Lohneysen2007,Stewart2001}. Non-Fermi liquid (NFL) behavior, complex magnetic order, charge order and unconventional superconductivity (SC) are examples of these states,
  which also have been observed in copper-oxide and iron pnictide high-\textit{T$_c$} superconductors, although often accompanied by electronic nematicity $-$ an electronic state that breaks 
  the rotational symmetry of the underlying lattice but not its translational symmetry \cite{Shibauchi2014, Fradkin2010, Fernandes2014}. The origin of the nematic state as well as its role 
  on the superconducting state
   remains controversial \cite{Lederer2017}. More recently, evidence for nematicity at high magnetic fields has been found in the heavy-fermion CeRhIn$_{5}$ \cite{Ronning2017}, indicating that
    superconductivity and nematicity also may be intertwined in this class of strongly correlated materials \cite{Okazaki2011}. Recent magnetostriction and nuclear magnetic resonance 
    experiments point to the importance of crystalline electric field (CEF) effects in the putative nematic state of CeRhIn$_{5}$ \cite{Rosa2019a,Lesseux2019}.

Ce$T$Sb$_{2}$ ($T =$ transition metal) is also a dense Kondo lattice system with pronounced CEF effects \cite{Muro1997,Thamizhavel2003}. The magnetic and CEF ground state of Ce$T$Sb$_{2}$ depend 
on the transition 
metal $T$ \cite{Thamizhavel2003,Andre2000}, and  CeAuSb$_{2}$ orders antiferromagnetically below $T_{N}= 5 - 6.8$~K depending on the occupancy of the Au site \cite{Balicas2005,Zhao2016,Seo2012}. 
Previous x-ray and neutron 
diffraction measurements showed that CeAuSb$_{2}$ crystallizes in a tetragonal crystal structure ($P4/nmm$) \cite{Flandorfer1996,Marcus2018}. Pressurizing CeAuSb$_{2}$ does not suppress $T_{N}$ to 
zero temperature, but rather induces new phases hindering the appearance of a QCP or SC \cite{Seo2012}. 
The nature of the pressure-induced phases in CeAuSb$_{2}$ remains an open question, but possibly stems 
from competing magnetic interactions known to exist in this series of compounds \cite{Adriano2014b}. Unlike pressure, magnetic fields are symmetry breaking and tend to localize $4f$ electrons. 
In CeAuSb$_{2}$, 
$T_{N}$ is gradually suppressed by magnetic fields applied along the $c$-axis and two metamagnetic (MM) transitions are observed at $H_{c1}$$ \sim 2.8$~T and $H_{c2}$$\sim5.6$~T at 
low temperature \cite{Balicas2005}. The 
$H$-$T$ phase diagram constructed by magnetoresistance measurements reveals that both MM transitions are first-order at low temperature and that $H_{c2}$ 
has a tricritical point at $3.7$~K \cite{Zhao2016}. In addition, a recent neutron diffraction study revealed that at $H_{c1}$ the magnetic structure changes from a single-$q$ striped phase with
 wave vector 
$\mathbf{Q}_{1}$=$(\eta, \eta, 1/2)$ [$\eta=0.136(2)]$ to a multi-$q$ (woven or checkered) phase with $\mathbf{Q}_{1}$=$(\eta, \eta, 1/2)$ and $\mathbf{Q}_{2}$=$(\eta, -\eta, 1/2)$ 
\cite{Marcus2018}. This ground state competition again stems from competing magnetic interactions, 
and resembles the single-$q$ to double-$q$ magnetic transition observed in pnictides \cite{Allred2016}. Further, striped phases appear to be ubiquitous in strongly correlated systems including manganites, cuprates, nickelates and cobaltites \cite{Mori1998,Tranquada1995,Chen1993,Vogt2000}.

Remarkably, both striped and woven magnetic phases in CeAuSb$_{2}$ have 2-fold rotational symmetry in the \textit{ab}-plane even though the underlying lattice has 4-fold rotational symmetry at high 
temperatures. As a result, one expects that the lattice will also break tetragonal \textit{C$_{4}$} symmetry in the presence of magnetoelastic coupling.
 In fact, in-plane uniaxial pressure experiments show that the magnetic transitions of CeAuSb$_{2}$ are sensitive to strain \cite{Park2018,Park2018a}. 
 The situation thus seems analogous to 
 iron-based high-$T_{c}$ superconductors such as NaFeAs (``111") and LaFeAsO (``1111"), which share the same $P4/nmm$ space group as CeAuSb$_{2}$ 
\cite{Lumsden2010,Chu2009}. There, the stripe magnetic order is generally preceded by a nematic phase with broken $C_{4}$ symmetry of the underlying lattice, in agreement with general 
theoretical expectations \cite{Xu2008a,Fang2008,Fernandes2012,Fernandes2019}.

Here we investigate whether such a nematic state exists in heavy-fermion CeAuSb$_{2}$.
Bulk thermodynamic probes show the presence of two transitions at about 6.5~K and 6.3~K in CeAuSb$_{2}$ at ambient pressure. X-ray diffraction measurements reveal that the development of 
zero-field striped magnetic order is connected to a structural transition. Our phenomenological model proposes that this structural transition must be a nematic transition at
 $T_{\mathrm{nem}} \geq T_{N}$. Our results support the scenario in which CeAuSb$_{2}$ hosts a nematic phase that is not intertwined with superconductivity.

\section{METHODS}

Single crystals of CeAuSb$_{2}$ were synthesized by a self-flux method described in Ref. \cite{Zhao2016}. The highest $T_{N}$ of 6.8 K is achieved in the electrical resistivity when the 
crystals are close to being stoichiometric, i.e., an average site occupancy of 100\% of Au and residual resistivity ratio (RRR) of about 20. Single crystals were pressurized to 2.66 GPa
 using a hybrid Be-Cu/NiCrAl clamp-type pressure cell. Daphne oil 7373 was used as a pressure-transmitting medium and lead was used as a manometer \cite{Eiling_1981}. 
 The in-plane electrical resistivity 
 of CeAuSb$_{2}$ was measured using a conventional four-probe technique with an LR700 Resistance Bridge in $^{4}$He and $^{3}$He cryostats from 300~K to 0.3~K. Magnetization measurements were 
 performed using a commercial superconducting quantum interference device (SQUID) and the specific heat was measured using a commercial small mass calorimeter that employs a quasi-adiabatic 
 thermal relaxation technique. Thermal expansion along the \textit{c}-axis was measured at atmospheric pressure using a capacitance cell dilatometer with a resolution in $\Delta L/L$ of 10$^{-8}$. 
 X-ray powder diffraction (XPD) measurements were performed at the X-ray diffraction and spectroscopy (XDS) beamline of the Brazilian Synchrotron Light Laboratory in Campinas, which uses 
 a 4~T superconducting multipolar wiggler source \cite{Lima2016}. The XPD patterns were collected at 20 keV in a transmission geometry using an area detector (MAR345). The sample was mounted at
  the cold finger of an open-cycle He cryostat (base temperature 2.3~K).

\section{RESULTS}

\begin{figure}[!ht]
\begin{center}
\hspace{-0.35cm}
\includegraphics[width=1\columnwidth]{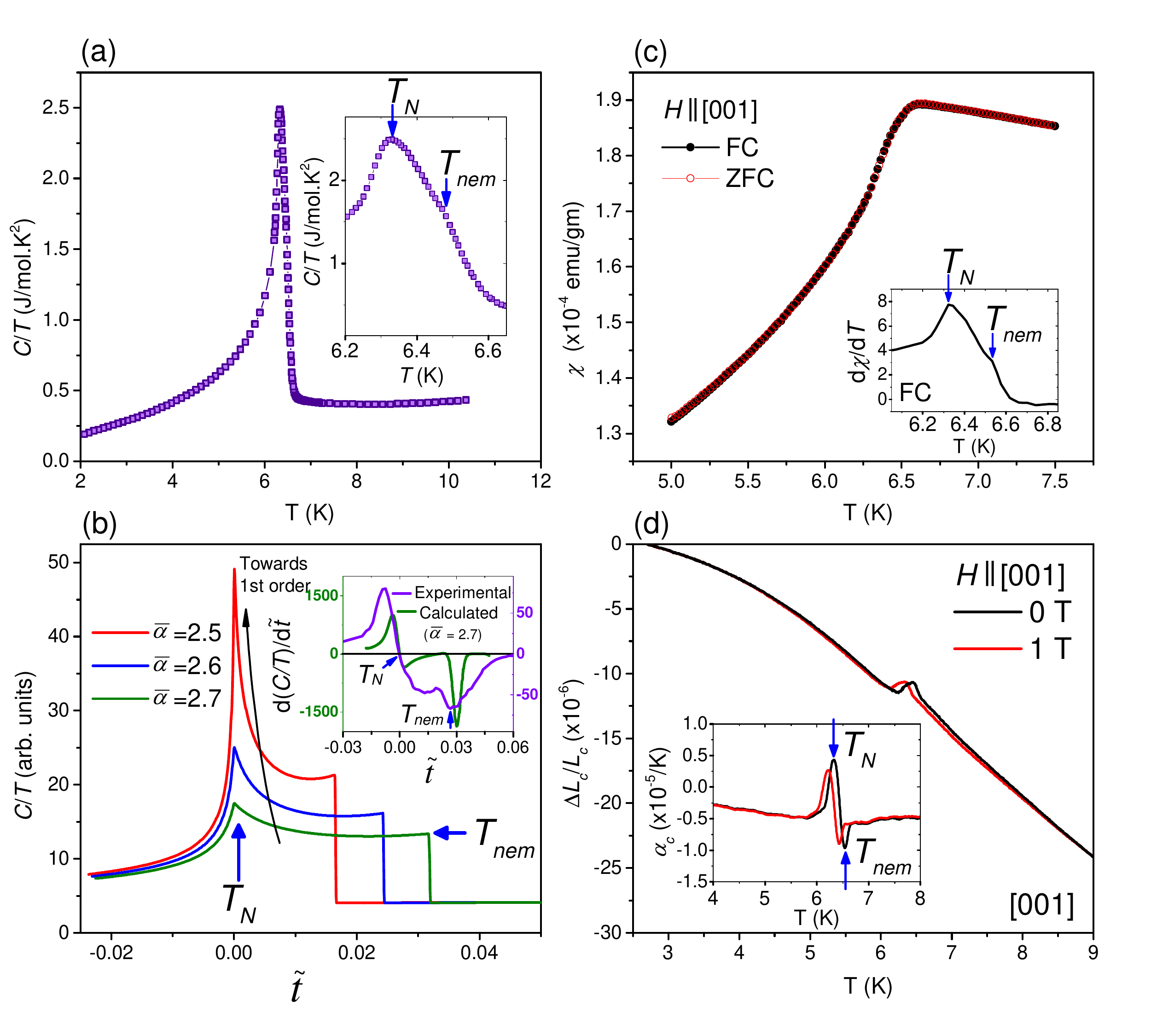}
\vspace{-0.65cm}
\end{center}
\caption{(a) Specific heat divided by temperature of CeAuSb$_{2}$ at ambient pressure.
 Inset is a magnified view near the peak position. Arrows indicate the transition temperatures $T_{\mathrm{nem}}$ and 
$T_{N}$. (b) Calculated $C/T$ versus reduced temperature for fixed anisotropy $\zeta$ 
and various inverse nematic coupling strength $\bar{\alpha}$. 
Inset gives the derivative of $C/T$ of the experimental data (purple) and of
the calculations with $\bar{\alpha} = 2.7$ (green). The theoretical curve was shifted and averaged over disorder for better comparison with 
the experimental data; details are given in the Supplementary Material. 
(c) Temperature dependent
magnetic susceptibility in an applied field of $1$~kOe along the $c$-axis. 
ZFC (zero-field cooled; open circles) and FC (field-cooled; solid circles) curves show two transitions without hysteresis. 
The inset plots the 
derivative d$\chi$/d$T$. 
(d) Temperature dependence of the thermal expansion $\Delta L/L$ of CeAuSb$_{2}$ along $[001]$ at zero field and 
1~T applied along the \textit{c}-axis. The inset gives the thermal expansion coefficient.} 
\label{fig:Fig1}
\end{figure}

Specific heat, magnetic susceptibility, and thermal expansion measurements in CeAuSb$_{2}$ reveal two closely lying transitions. As shown in Fig. 1(a), and highlighted in the inset of Fig.~1(b), 
two peaks occur in the specific heat at 
$T_{\mathrm{nem}} = 6.48$~K and $T_{N} = 6.33$~K, indicated by arrows. Figure 1(c) shows the temperature dependence of the magnetic susceptibility for $H || [001]$, and the presence of two 
distinct phase transitions is evident in the derivative d$\chi/dT$ at $T_{\mathrm{nem}} = 6.53$~K and $T_{N} = 6.33$~K shown in the inset. Figure 1(d) shows the temperature dependence of the 
\textit{c}-axis length change, $\Delta L_{c}/L_{c}$, with an anomaly in the vicinity of 6.5~K. This anomaly is not typical of purely magnetic phase transitions \cite{Takeuchi2001,Takeuchi2003}, and the
 linear thermal-expansion 
coefficient, $\alpha_{c} = (1/L_{c})(dL_{c}/dT)$, further reveals a negative peak at $T_{\mathrm{nem}} = 6.55$~K as well as a positive peak at $T_{N} = 6.33$~K, respectively (see inset of Fig. 1(d)). 
$T_{\mathrm{nem}}$ and $T_{N}$ decrease together when a field of 1~T is applied, suggesting a strong coupling of the order parameters associated with these transitions. As we will come 
to below, $T_{\mathrm{nem}}$ is associated with a structural transition which likely breaks $C_{4}$ symmetry and $T_{N}$ with a magnetic transition to the single-$q$ stripe phase.

\begin{figure*}[!ht]
\begin{center}
\includegraphics[width=1.5\columnwidth,keepaspectratio]{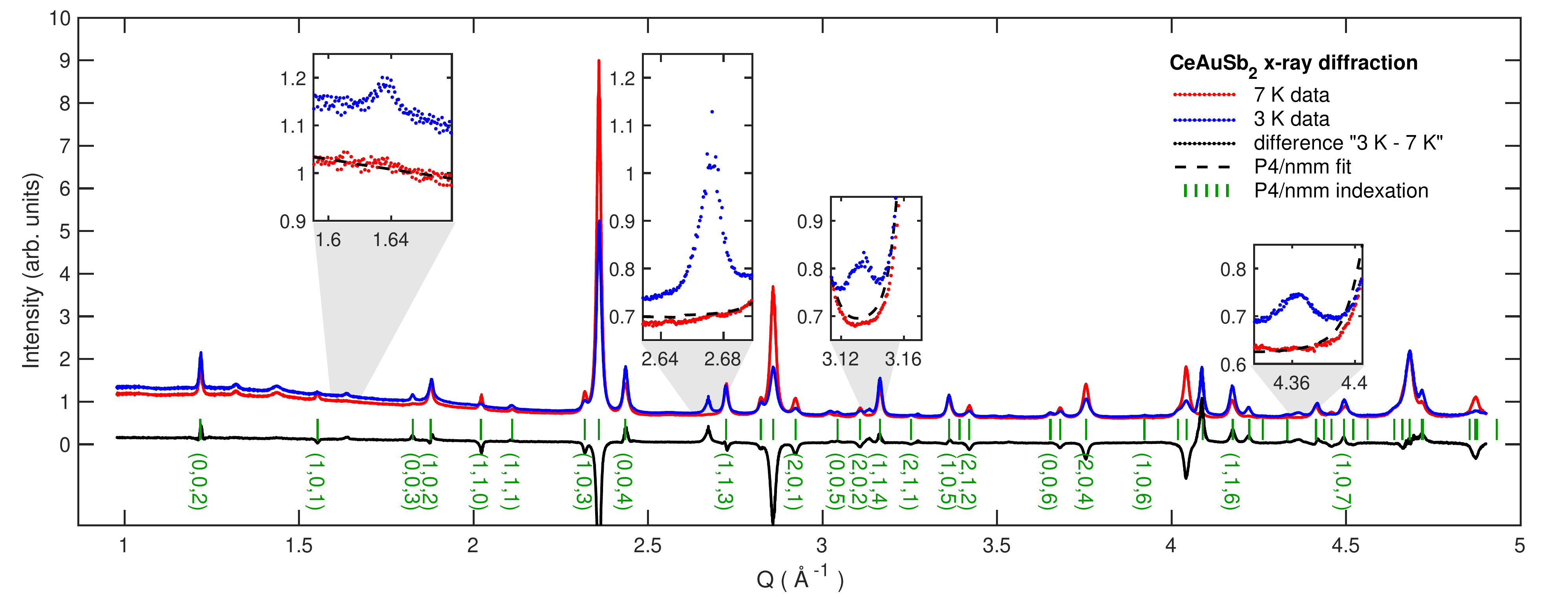}
\vspace{-0.5cm}
\end{center}
\caption{Synchrotron (20 keV) x-ray powder diffraction patterns of CeAuSb$_{2}$ obtained above ($7$~K, red) and below ($3$~K, blue) the transitions at $T_{N,\mathrm{nem}}$. 
The difference pattern (black line) emphasizes significant intensity gains, or losses, at most Bragg positions. Green markers and labels indicate the indexation of these peaks in the high-temperature $P4/nmm$ structure.
 A lattice distortion is not resolved. Below the structural transition, four additional Bragg peaks are observed (see insets), which are not indexed by the parent phase.}
\label{fig:Fig2}
\end{figure*}

X-ray powder diffraction patterns of CeAuSb$_{2}$ measured at $T$=$7$~K and $T$=$3$~K are shown in Fig.~2. Rietveld refinement of the 7 K data in the published structure \cite{Sologub1994}
provides a satisfactory fit in the $P4/nmm$ cell with lattice parameters $a = 4.3954(2)$\AA$\,$ and $c = 10.318(1)$\AA. As indicated by the black difference 
line ($I(3$K$) - I(7 $K)) in Fig.~2, the intensities of most Bragg peaks change (i.e., increase or decrease) below $T_{N}/T_{\mathrm{nem}}$, whereas a number of intensities also appear unaffected. 
Within the limited resolution of this experiment ($\Delta d/d > 10^{-4}$), we observe no evidence for a lattice distortion. As illustrated by the insets in Fig. 2, 
however, four peaks appear at low temperatures which are not indexed in the parent phase and thus confirm that the true symmetry of the low temperature phase of CeAuSb$_{2}$ is lower than that of the $P4/nmm$ space group. Due to the powder average and the limited number of observed intensities in the present dataset, the full solution of the low-temperature modulated structure will have to await
 a single crystal diffraction study. Nevertheless, the absence of nuclear satellite peaks in neutron diffraction measurements \cite{Marcus2018} suggests that the x-ray diffraction superstructure peaks are due to a modulation of charge density, which is intimately coupled to the magnetic order parameter.

\begin{figure*}[!ht]
\begin{center}
\includegraphics[width=1.25\columnwidth,keepaspectratio]{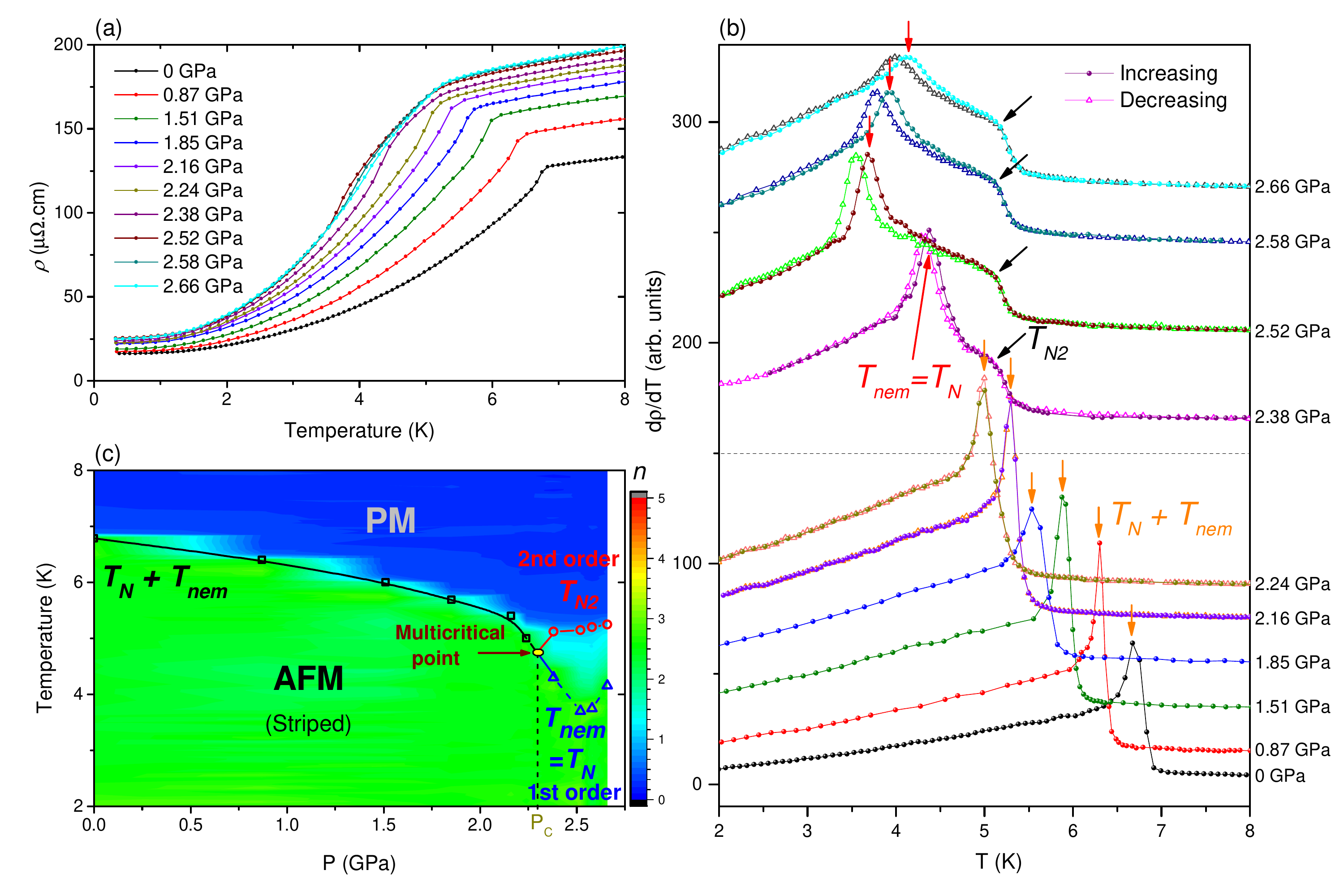}
\vspace{-0.5cm}
\end{center}
\caption{(a) Temperature dependence of the in-plane electrical resistivity, $\rho_{ab}$, for CeAuSb$_{2}$ at pressures to 2.66 GPa. 
(b) Temperature derivative of $\rho_{ab}$ under pressure. Data at different pressures are shifted for clarity. 
Arrows indicate peak positions: orange, black, and red arrows for the 
$T_{N} + T_{\mathrm{nem}}$, $T_{N2}$, and $T_{\mathrm{nem}} = T_{N}$, respectively. 
Above 2.16~GPa, the results from increasing- and decreasing-temperature ramps are indicated by solid and open symbols, 
respectively. (c) $T$-$P$ phase diagram of CeAuSb$_{2}$ in zero applied field. 
 Colors represent the local exponent, $n=\partial \mathrm{ln} \Delta \rho / \partial \mathrm{ln} T$.}
\label{fig:Fig3}
\end{figure*}

The existence of a magnetic ordered state that breaks a point-group symmetry (i.e., tetragonal symmetry), besides time-reversal symmetry, suggests that the phase transition can happen in two 
stages \cite{Fernandes2019}: at $T_{\mathrm{nem}}$, tetragonal symmetry is broken, whereas at $T_{N}$, time-reversal symmetry is broken. In order to describe these transitions in CeAuSb$_{2}$, 
we consider a phenomenological model using a stripe magnetic order with a two-component order parameter $\mathbf{m} = (\mathbf{m_{1}}, \mathbf{m_{2}})$, with ordering wavevectors
$\mathbf{Q}_{1} = (\eta, \eta, 1/2)$ and $\mathbf{Q}_{2} = (\eta, -\eta, 1/2)$  for $\mathbf{m_{1}}$ and $\mathbf{m_{2}}$, respectively. Note that these two ordering vectors are not equivalent 
in the
 $P4/nmm$ group, but are related by a 90$^{\circ}$ rotation. Up to quartic terms, the Ginzburg-Landau free energy compatible with the symmetries present in CeAuSb$_{2}$ is given by:

\begin{equation}
F = \frac{1}{2}r_{0}(\mathbf{m}_{1}^{2}+\mathbf{m}_{2}^{2})+ \frac{u}{4} (\mathbf{m}_{1}^{2}+\mathbf{m}_{2}^{2})^{2} - \frac{g}{4} (\mathbf{m}_{1}^{2}-\mathbf{m}_{2}^{2})^{2},
\end{equation}

\noindent where $r_{0} \propto T - T_{N}^{0}$, $u > g >0$.  This is identical to the free energy employed to understand the nematic phase of the pnictides \cite{Fernandes2012}. Here for simplicity
 we have considered both $\mathbf{m_{1}}$ and $\mathbf{m_{2}}$ to be real, and that contributions from higher harmonics are neglected. Within a mean-field approximation, at $T < T_{N}^{0}$, 
 the free energy is minimized by developing a stripe magnetic order, with either $\langle\mathbf{m_{1}}\rangle$ or 
$\langle\mathbf{m_{2}}\rangle$  becoming non-zero. Such a magnetic order also breaks $C_{4}$ symmetry down to $C_{2}$. Therefore, $T_{N}^{0}$ marks a simultaneous second-order nematic and 
magnetic phase transition. Going 
beyond mean field analysis and including the effects of magnetic fluctuations, this model generally predicts a nematic transition at $T_{\mathrm{nem}} \geq T_{N}$, 
where $T_{N}$ is the renormalized magnetic transition temperature to the stripe phase \cite{Fernandes2012}. Such a paramagnetic-nematic state is characterized by anisotropic magnetic fluctuations, 
$\langle\mathbf{m_{1}}^{2}-\mathbf{m_{2}}^{2}\rangle \neq 0$, which spontaneously break the $C_{4}$ symmetry of the system even 
though $\langle\mathbf{m_{1}}\rangle = \langle\mathbf{m_{2}}\rangle = 0$. The fate of the 
coupled magnetic-nematic transitions depends on the parameters of the phenomenological model, which in our case are the degree of anisotropy of the magnetic fluctuations and the 
nematic coupling strength. The former is 
given by the ratio $\zeta \equiv J_{z}/J$ between the out-of-plane and in-plane effective exchange interactions, whereas the latter is given by the ratio $\bar{\alpha} \equiv u/g$ of the 
coefficients of Eq. (1). 
As shown previously in Ref.~\cite{Fernandes2012}, for a fixed 
$\zeta$, large $\bar{\alpha}$ gives two split second-order transitions at $T_{\mathrm{nem}} > T_{N}$, whereas small $\bar{\alpha}$ gives a simultaneous first-order transition. 
In the intermediate $\bar{\alpha}$ range, the transitions 
remain split, but one of them becomes first-order \cite{Fernandes2012} (see Supplementary Material for details).  We computed the entropy and heat capacity for a fixed $\zeta$ and 
decreasing $\bar{\alpha}$ towards the critical value 
$\bar{\alpha}_{c}$ below which the magnetic transition becomes first-order but remains split from the nematic transition (see Supplementary Materials for details). Figure 1(b) 
shows the calculated heat capacity $C/T$ as a 
function of reduced temperature 
$\tilde{t}\propto ({T-T_{N}})/{T_{N}}$. 
Upon approaching $\bar{\alpha}_{c}$, the specific heat peak at $T_{N}$ increases at a faster rate than the jump at $T_{\mathrm{nem}}$, and the separation between the 
two transition temperatures decreases, signaling the approach of a first-order magnetic transition. In order to compare our model with the experimental result in Fig 1(a), which do not show sharp jumps, we include 
weak disorder by considering a distribution of $T_{N}^{0}$ values, yielding a smooth temperature dependence of the derivative of $C/T$ (see inset of Fig. 1(b)). The comparison shows that d($C/T$)/d$\tilde{t}$ = 0 at $T_{N}$ and d($C/T$)/d$\tilde{t}$ has a minimum 
at $T_{\mathrm{nem}}$ and suggests 
that CeAuSb$_{2}$ at ambient conditions is close to $\bar{\alpha}=2.7$. Our model predicts that below $\bar{\alpha}_{c}$, both nematic and magnetic phase transitions become 
first-order and simultaneous at 
$T_{\mathrm{nem}}=T_{N}$. As we will come to below, our experimental data under applied pressure suggest that hydrostatic pressure might reduce $\bar{\alpha}$, thus driving the
 transition first-order and simultaneous.
 
Now we turn our attention to the field and pressure dependence of the coupled transitions in CeAuSb$_{2}$. Figure 3(a) shows the temperature dependence of the in-plane 
resistivity ($\rho_{ab}$) of CeAuSb$_{2}$ under 
various pressures up to 2.66 GPa. At ambient pressure, a sharp drop in $\rho_{ab}$ occurs at $\sim 6.8$~K. As shown in Fig. 3(b), at low pressure a single peak marked
 by orange arrows occurs in d$\rho$/d$T$. 
Within the resolution 
of these data ($\Delta T \sim 0.08$~K), this single peak likely encloses both $T_{N}$ and $T_{\mathrm{nem}}$ found in $C/T$, d$\chi$/d$T$, and $\Delta L/L$. With increasing 
pressure, $T_{N} + T_{\mathrm{nem}}$ is 
suppressed to 5 K at 2.24 GPa. Above 2.24 GPa, however, the single peak splits into two peaks in d$\rho$/d\textit{T} which are indicated by black and red arrows, respectively. $T_{N2}$ marks
 a shoulder-like broad peak which 
increases with increasing pressure and $T_{\mathrm{nem}} = T_{N}$ marks a sharp symmetric peak which decreases up to 2.52 GPa but increases above 2.52 GPa. We will discuss the origin 
of these transitions below. 
Above 2.16 GPa, the results from increasing- and decreasing-temperature ramps are shown together in Fig. 3(b), which are indicated by solid and open symbols, respectively. Interestingly, 
a hysteresis appears at 
$T_{\mathrm{nem}}=T_{N}$ in d$\rho$/d$T$, pointing to a first-order phase transition. Hysteresis is not typical of a ‘naked’ antiferromagnetic (AFM) transition but rather suggests the 
development of a multi component state. 
At $T_{N2}$ and $T_{N} + T_{\mathrm{nem}}$, however, hysteresis is not observed in the resistivity. To shed light on the temperature-pressure phase diagram shown in Fig. 3(c), we turn to the
 pressure dependence of the initially coupled transitions estimated \textit{via} Ehrenfest relation:

\begin{equation}
\frac{dT_{N/ \mathrm{nem}}}{dP} = \frac{V_{m}\Delta\beta}{\Delta C/T}
\end{equation}

\noindent where $V_{m}$ is the molar volume, $\Delta\beta$ and $\Delta C/T$ are changes of the volume thermal expansion coefficient and the specific heat divided by temperature at the 
transition. By using the experimental values of $\beta$ and $C/T$ at ambient pressure, we obtain d$T_{N}$/d$P = 0.37$~K/GPa 
and d$T_{\mathrm{nem}}$/d$P = - 0.77$~K/GPa (see supplementary Fig. S3), i.e., $T_{\mathrm{nem}}$ would be suppressed 
with pressure, whereas $T_{N}$ would be enhanced with pressure.
These transitions, however, are coupled at low pressures, and the sum of the calculated d\textit{T}/d\textit{P} values ($-$0.4 K/GPa) is in good agreement with the experimental pressure dependence of coupled 
transition, 
d($T_{N} + T_{\mathrm{nem}}$)/d$P \approx - 0.45$~K/GPa. The fact that $T_{N2}$ increases with pressure above the critical pressure suggests that it is associated with a different ordered state not coupled to a tetragonal-symmetry breaking, whose origin will be discussed below.


\begin{figure*}[!ht]
\begin{center}
\includegraphics[width=1.5\columnwidth,keepaspectratio]{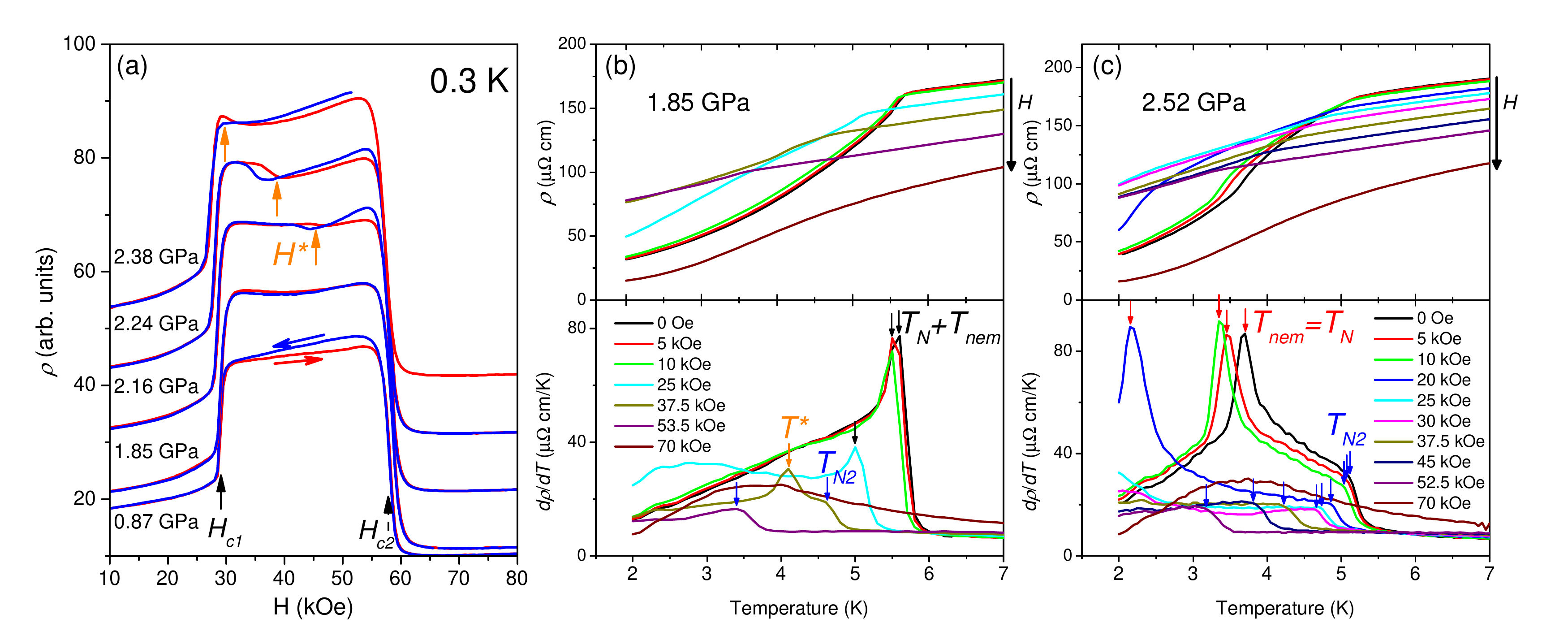}
\end{center}
\vspace{-0.5cm}
\caption{(a) In-plane electrical resistivity of CeAuSb$_{2}$ as a function of magnetic field applied along 
the \textit{c}-axis at 0.3 K for representative pressures. Magnetoresistance from increasing- and decreasing-field ramps are 
plotted by red and blue lines, respectively. Data at different pressures are shifted for clarity. 
(b), (c) Low-temperature in-plane electrical resistivity for various applied magnetic fields at 1.85 GPa and 2.52 GPa. 
$\rho_{ab}$ and d$\rho$/d$T$ as a function of temperature are shown in upper and bottom panels, respectively. }
\label{fig:Fig4}
\end{figure*}

The temperature-pressure false-color map of the local exponent \textit{n} of the temperature dependence of $\rho_{ab}$ is shown in Fig. 3(c), 
where $n=\partial \mathrm{ln} \Delta \rho / \partial \mathrm{ln} T$ and $\Delta \rho = \rho_{ab} - \rho_{0} = AT^{n}$.  The contour map in Fig. 3(c) provides a proxy for the magnetic scattering in 
different regions of the phase diagram. Above $P_{c}$, the green colored region ($n \approx 2$) below $T_{\mathrm{nem}} = T_{N}$ displays the same behavior as the zero-pressure phase, 
indicating that the magnetic phase is striped AFM, whereas below 
$T_{N2}$ the light blue colored region indicates a distinct behavior ($n \approx 1.2$). We will show below in Fig. 5 that the power-law behavior in light blue colored region at high 
pressure is similar to the behavior 
of the multi-$q$ AFM phase at ambient pressure and high fields. Therefore, our results suggest that the coupled transition at $T_{N} + T_{\mathrm{nem}}$ decouples at $P_{c}$ into (1) 
a second-order AFM transition to the multi-$q$ phase at 
$T_{N2}$ and (2) a simultaneous first-order structural and magnetic phase transition to the stripe phase at $T_{\mathrm{nem}}=T_{N}$ due to a finite magnetoelastic coupling. 
This suggests the existence of a multicritical point at $P_{c}$ indicated by a yellow circle in Fig. 3(c) at which four phases meet: disordered phase ‘PM’, 
paramagnetic-nematic phase ‘PMnem’, antiferromagnetically ordered stripe-nematic phase ‘AFM(S)nem’, and antiferromagnetically ordered multi-$q$ phase ‘AFM(M)’.

Evidence for a hysteretic first-order transition is further confirmed in Fig. 4(a), which shows the field dependence of 
the in-plane resistivity under various pressures, for increasing- and 
decreasing-field ramps applied along the $c$-axis. 
At 0.3 K, the first metamagnetic (MM) transition field $H_{c1}$ and the second MM transition field $H_{c2}$ do not change with increasing pressure as shown in Fig. 4(a). At low 
pressure, the magnetoresistance between $H_{c1}$ and $H_{c2}$ monotonically increases with increasing field, in agreement with Ref. \cite{Zhao2016}. Above 2.16 GPa, however, a hysteretic 
step-like anomaly at $H^{*}$ is observed between $H_{c1}$ and $H_{c2}$. 
$H^{*}$ decreases with increasing pressure and temperature (see supplementary Fig. S4). Figures 4(b) and (c) show the temperature dependence of $\rho_{ab}$ and d$\rho$/d$T$ at 
representative pressures for CeAuSb$_{2}$ in an applied magnetic field. At 1.85 GPa, the coupled transition at $T_{N} + T_{\mathrm{nem}}$ turns into two transitions 
at $T^{*}$ and $T_{N2}$ above $H_{c1}$. The new field-induced phase transition at
 $T^{*}$ ($H>H_{c1}$) matches the transition at $H^{*}$ (see supplementary Fig. S7). 

\begin{figure}[b]
\begin{center}
\includegraphics[width=1\columnwidth,keepaspectratio]{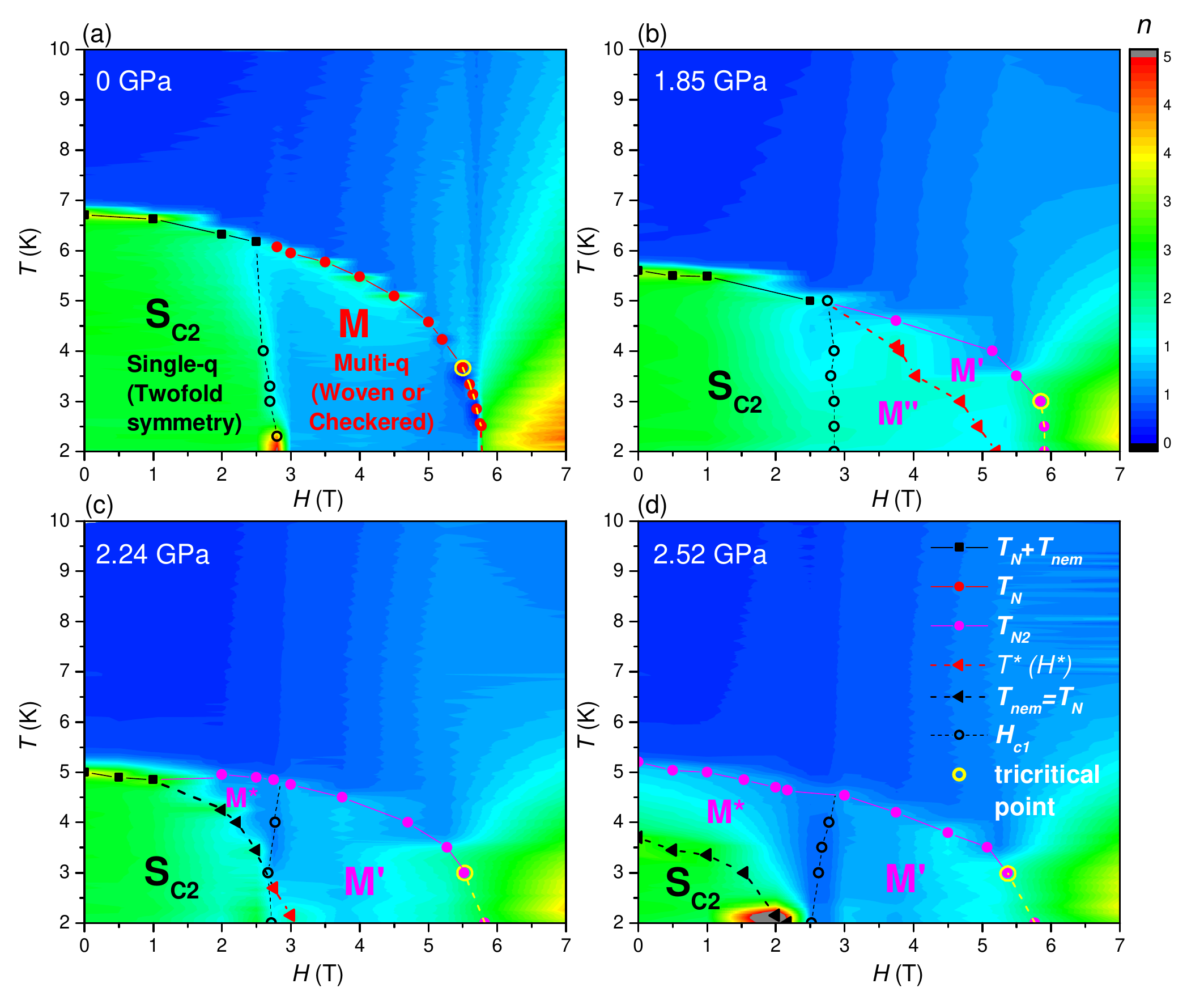}
\end{center}
\vspace{-0.5cm}
\caption{False-color map of the effective resistivity exponent $n$ in the $H$-$T$ phase diagram at (a) 0 GPa, (b) 1.85 GPa,
 (c) 2.24 GPa, and (d) 2.52 GPa. The green region denotes the single-$q$ stripe phase (S$_{C2}$) and the region of light blue 
 denotes the other possible AFM phases (M, M', M'', and M*).}
\label{fig:Fig5}
\end{figure}

Figure 5 displays the temperature-magnetic field false-color map of the local exponent $n$ of $\rho_{ab}$ for representative pressures, 
where $n=\partial \mathrm{ln} \Delta \rho / \partial \mathrm{ln} T$
and $\Delta \rho = \rho_{ab} - \rho_{0} = AT^{n}$. The contour maps show the presence of two distinct regions in the AFM phase diagrams. The single-$q$ region 
(S) displays a local exponent $n \approx 2$, whereas the multi-$q$ region (M) reveals $n \approx 1$ at ambient pressure. At 2.52 GPa, the contour map indicates that 
 the pressure-induced new phase in zero field below $T_{N2}$ is related to the field-driven multi-$q$ phase at ambient pressure. In order to understand the origin 
of $T^{*}$ and the other phases denoted by (M', M'', and M*) under pressure, the underlying magnetic structures will have to be determined by neutron or nuclear
magnetic resonance measurements under applied pressure and field. Interestingly, experiments on CeAuSb$_{2}$ under uniaxial strain along the $[100]$ direction 
also suggest additional magnetic phases \cite{Park2018,Park2018a}. Finally, we note that the tricritical point of $H_{c2}$ is not suppressed to zero temperature with 
pressure (see supplementary information). 

\section{DISCUSSION}

Striped magnetic phases are ubiquitous in strongly correlated materials  \cite{Mori1998,Vogt2000,Khalyavin2008,Knizek1992}, 
and unconventional superconductivity is arguably intertwined with spin and charge stripe correlations in the copper oxides and iron-based materials \cite{Tranquada1995}.  Superconductivity, 
however, is absent in e.g. nickel oxides,  which also host a stripe pattern within the NiO$_{2}$ planes analogous to the CuO$_{2}$ planes in copper oxides \cite{Chen1993}. Here we show that the 
$f$-electron system CeAuSb$_{2}$ also supports a nematic state in the absence of superconductivity. Though nematicity may boost Cooper pairing in the iron-based superconductors, 
nematic fluctuations may be weak in CeAuSb$_{2}$ due to the first-order nature of the nematic transition and the change in the ground-state wavefunction of the system at $P_{c}$.  
Previous pressure work revealed that two energy scales, one associated with Kondo coherence and the other with crystalline electric field (CEF) splitting become similar at $P_{c}$.
As a result, wave functions of the excited crystal-field levels become admixed into the ground state, hindering formation of a fully Kondo coherent ground state yet allowing new
 magnetic orders which lead to a magnetic state that persists to over 4 GPa \cite{Seo2012}. Because of the competing magnetic interactions known to be present in the class of localized materials, 
 various magnetic orders compete under pressure and magnetic field. The family member CeAgSb$_{2}$ also shows a new magnetic phase above 
$P_{c} \sim 2.7$~GPa and $T_{max}$ exhibits unexpected negative pressure dependence above $P_{c}$ which may be due to the influence of a low-lying CEF level $\Delta_{1} \sim 50$~K  
\cite{Takeuchi2003,Sidorov2003}. Inelastic and elastic neutron scattering studies could shed light on the relation between new magnetic phases and CEF admixture under applied pressure 
and magnetic field. 

\section{CONCLUSIONS}

In summary, we have constructed the temperature-magnetic field phase diagrams of CeAuSb$_{2}$ under applied pressure. Bulk thermodynamic probes reveal two closely lying phase transitions at $T_{N} = 6.3$~K 
and $T_{\mathrm{nem}}= 6.5$~K, and x-ray diffraction measurements verify that striped magnetic order is connected to a structural phase transition at ambient pressure. Our theoretical model suggests that stripe magnetic order is preceded by a nematic phase in CeAuSb$_{2}$. The discovery of a putative nematic phase in an $f$-electron material at zero magnetic field provides an unexplored framework for nematicity, though superconductivity – if present – remains to be discovered.

\begin{acknowledgments}
Work at Los Alamos was performed under the auspices of the U.S. Department of Energy, Office of Basic Energy Sciences, Division of Materials Science and Engineering. S.S. acknowledges a Director’s Postdoctoral Fellowship through the Laboratory Directed Research \& Development program. Scanning electron microscope and energy dispersive X-ray measurements were performed at the Center for Integrated Nanotechnologies, an Office of Science User Facility operated for the U.S. Department of Energy (DOE) Office of Science. Theory work (R.M.F.) was supported by the U.S. Department of Energy, Office of Science, Basic Energy Sciences, under Award No. DE-SC0012336. XW is supported by the University of Chicago Materials Research Science and Engineering Center, which is funded by the National Science Foundation under award number DMR-1420709. R.D. dos Reis and D. Carmo thanks the financial support from the Brazilian agencies CAPES and FAPESP (Grants 2018/00823-0 and 2018/22883-5).

\end{acknowledgments}

%

\end{document}